\documentclass{aa}
\usepackage{graphicx}      
\usepackage{txfonts}
\usepackage{natbib}
\usepackage{epsfig}
\usepackage{subfigure}
\bibpunct{(}{)}{;}{a}{}{,} 

\begin{document}
   \title{Radiative and mechanical feedback into the molecular gas of \\
   NGC 253}

   \author{M.~J.~F.~Rosenberg
          \inst{1}
          \and
          M.~V. Kazandjian \inst{1}
          \and
          P.~P. van der Werf \inst{1}
          \and
          F.~P. Israel \inst{1}
          \and
          R. Meijerink \inst{2}
          \and
          A. {Wei{\ss}} \inst{3}
          \and
          M.~A. Requena-Torres \inst{3}
          \and
          R. G{\"u}sten \inst{3}
          }

  \institute{Sterrewacht Leiden, Universiteit Leiden,
              P.O. Box 9513, NL-2300 RA Leiden\\
              \email{rosenberg@strw.leidenuniv.nl}
              \and
             Kapteyn Astronomical Institute, University of Groningen,\\
             P.O. Box 800, NL-9700 AV Groningen, The Netherlands
              \and 
             Max-Planck-Institut für Radioastronomie,\\
             Auf dem Hügel 16, Bonn, D-53121, Germany
             }

   \date{Accepted: January 20, 2014}

\abstract{Starburst galaxies are undergoing intense periods of star formation.  Understanding 
the heating and cooling mechanisms in these galaxies can give us insight to the driving mechanisms that fuel 
the starburst.  Molecular emission lines play a crucial role in the cooling of the excited gas. With SPIRE on the Herschel Space 
Observatory we have observed the rich molecular spectrum towards the central region of NGC 253. CO transitions from J=4-3 to 13-12 are observed 
and together with low-J line fluxes from ground based observations, these lines trace the excitation of CO. 
By studying the CO excitation ladder and comparing the intensities to models, we investigate whether the 
gas is excited by UV radiation, X-rays, cosmic rays, or turbulent heating.  Comparing the $^{12}$CO and $^{13}$CO observations to large velocity gradient models and PDR models we find three main ISM phases.
We estimate the density, temperature,and masses of these ISM phases.  By adding $^{13}$CO, HCN, and HNC line intensities, 
we are able to constrain these degeneracies and determine the heating sources.
The first ISM phase responsible for the low-J CO lines is excited by PDRs, but the second and third phases, responsible for the mid to high-J CO transitions, 
require an additional heating source.
We find three possible combinations of models that can reproduce our observed molecular emission.  Although we cannot
determine which of these are preferable, we can conclude that mechanical 
heating is necessary to reproduce the observed molecular emission and cosmic ray heating is a negligible heating source.  We then estimate the mass of each ISM phase; 6$\times 10^7$ M$_\odot$ for phase 1 (low-J CO lines), 
3$\times 10^7$ M$_\odot$ for phase 2 (mid-J CO lines), and 9$\times 10^6$ M$_\odot$ for phase 3 (high-J CO lines) for a total system mass of 1$\times10^{8}$ M$_\odot$.}

\titlerunning{Molecular gas excitation in NGC 253}
\authorrunning{Rosenberg, M.~J.~F. et al.}

\maketitle

\section{Introduction}

Starburst galaxies are nearby laboratories that allow us to study intense star formation. Studying the heating and cooling mechanisms in these galaxies gives us insight 
in to which excitation or feedback mechanisms are dominant in fueling starbursts.  Molecular emission lines play a crucial role in the cooling of excited gas, and 
with the Herschel Space Observatory we are able to observe the rich molecular spectrum of the nucleus of NGC 253.  

CO is one of the most abundant molecules after H$_2$, and a good and easily observable tracer of the
condition of the molecular gas in the interstellar medium (ISM) of these galaxies.  CO in the ISM is mostly located in molecular clouds and photon dominated regions (PDRs), where
the radiation can penetrate the cloud and excite the gas.  \citet{1985ApJ...291..722T} created models of PDRs, which predict the intensities of atoms and molecules in the PDR as a function of
density, radiation environment, and column density.  These have been expanded to include models for X-ray dominated regions (XDRs)\citep{2005A&A...436..397M,1996ApJ...466..561M}, PDRs with enhanced cosmic 
ray ionization rates \citep{2006ApJ...650L.103M}, and PDRs with additional mechanical heating taken into account (Kazandjian et. al, in press). 

NGC 253 is a nearby, D$_L$=2.5 Mpc (\cite{1990AJ....100..102D}, 12 pc/''), edge-on barred spiral galaxy \citep{1985ApJ...289..129S}.  The central kiloparsec of NGC 253
is considered an archetypal starburst nucleus ($L_{IR}\sim2\times10^{10}$ L$_\odot$), which is heavily obscured at optical wavelengths by dust lanes 
\citep{1996ApJ...458..537P}.  However, in the far-infrared and submillimeter wavelength regimes, NGC 253 exhibits extremely
bright molecular line transitions \citep{1991A&ARv...3...47H}, originating from large molecular clouds in the nuclear region \citep{1995A&A...302..343I,1996A&A...305..421M,1997A&A...325..923H,
2003ApJ...586..891B,2009ApJ...706.1323M}.  This gas also appears to be highly excited.  Observations of HCO$^+$ and HCN suggest
that at least some of the gas has densities greater than 10$^4$ cm$^{-3}$ and temperatures over 100 K \citep{1995ApJ...454L.117P,1997ApJ...484..656P}.  

There have been many studies that have attempted to derive the excitation mechanism in NGC 253.  The warm, excited molecular gas phase, 
often associated with PDRs, excites the mid- to high-J 
CO transitions ($J>4$).  The near infrared H$_2$ emission lines shows that PDRs are an important excitation mechanism \citep{2013A&A...550A..12R}. However hot H$_2$ gas only traces
the very edges of molecular clouds.  In order to study the excitation of the bulk of the molecular gas, we may use CO as a probe.
\citet{2003ApJ...586..891B} have observed $^{12}$CO up to J=6-5 along with $^{13}$CO up to J=3-2 and derive a kinetic
temperature of 120 K and an H$_2$ density of 4.5$\times10^4$ cm$^{-3}$ for the warm phase.  However, they 
suggest that the CO is excited by cosmic rays, and not only by PDRs.  \citet{2006ApJS..164..450M} found that the chemistry and
heating of NGC 253 is dominated by large scale, low velocity shocks.  Presence of shocked molecular material is 
indeed evident through the presence of widespread SiO emission throughout the nuclear region \citep{2000A&A...355..499G}.  In addition,
\citet{2008ApJ...689L.109H} detected the first extragalactic $^{13}$CO J=6-5 transition, and using this determined that shocks are the dominant excitation mechanism
in the nuclear region of NGC 253. 
\citet{2009ApJ...706.1323M} suggest that, although NGC 253 is dominated by shock chemistry, PDRs also play a crucial
role in the chemistry, since there are very high abundances of PDR tracing molecules, namely HCO$^+$, CO$^+$.

In order to better constrain the dominant excitation mechanism in NGC 253, we present the full $^{12}$CO ladder
up to J=13-12, the $^{13}$CO up to J=6-5 as observed with the Herschel Space Observatory. We combine this with observations of HNC, and HCN transitions 
and apply these observations to models of PDRs, XDRs, enhanced cosmic ray PDRs, and 
enhanced mechanical heating PDRs in order to model the excitation directly.  These observations were taken as part of the Guaranteed Key Program Herschel EXtra GALactic
(HEXGAL, PI: R. G{\"u}sten). In Section~\ref{sec:obs}, we will describe the observations and data reduction techniques.
In Section~\ref{sec:res}, we will present our spectra and line fluxes.  Using models of CO emission, 
in Section~\ref{sec:coladder} we will constrain the parameters of the molecular gas phases and introduce a methodology
to understand the degeneracies of the models.  In order to constrain the densest phase of the ISM,
we use the HCO$^+$ and HCN in Section~\ref{sec:highj} to determine the excitation mechanism.   In Section~\ref{sec:disc}
we analyze the implication of our results and in Section~\ref{sec:conc} we summarize our main conclusion.

\section{Observations and Data Reduction}
\label{sec:obs}

Observations of NGC 253 were taken on December 5th, 2010 with the Herschel Spectral and Photometric Imaging Receiver (SPIRE) in staring mode centered on the nucleus of NGC 253 
(Obs ID: 1342210847).  SPIRE is an imaging Fourier Transform Spectrometer (FTS) \citep{2010A&A...518L...3G}.
The high spectral resolution mode was used with a resolution of 1.2 GHz over both observing bands.  
The low frequency band covers $\nu$=447-989 GHz ($\lambda$=671-303 $\mu$m) and the high frequency band 
covers $\nu$-958-1545 GHz ($\lambda$=313-194 $\mu$m).  A reference measurement was used to subtract the emission from the telescope and instrument.  
%

\begin{figure}
\resizebox{\hsize}{!}{\includegraphics{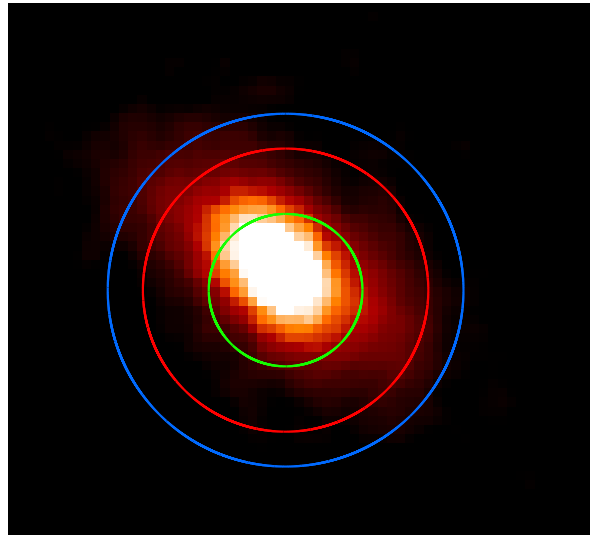}}
\caption{Three SPIRE beam sizes are shown overplotted on a SCUBA 450 $\mu$m archival image,  40.5'' (blue), 17.5'' (green), and 32.5'' (red). The SPIRE beamsize 
changes as a function of wavelength, thus the blue circle represents the largest beam size and the green represents the smallest beam size.
The red beam size (32.5'') represents the beam size of CO J= 5-4 transition and all other lines are convolved to this beam size.}
\label{fig:ngc253}
\end{figure}

\label{sec:obs1}
The data were reduced using version 9.0 of the Herschel Interactive Processing Environment (HIPE). Since NGC 253 
is an extended source, a beam correction factor is necessary to compensate for the wavelength dependent beam 
size. Using an archival SCUBA 450 $\mu$m map of NGC 253, we convolve the map with a 2-D Gaussian with FWHM the 
same size as our beam sizes.  The archival 450 $\mu$m SCUBA map of NGC 253 is shown in Figure\ref{fig:ngc253} with the largest, smallest, and 
normalized beam sizes are shown in blue, green, and red respectively.  All beam sizes include the brightest part 
of the nucleus.

The ratio of the flux within each convolved map and the flux within the beam size of the CO J=5-4 
transition (beam size=32.5'') is the beam correction factor ($\kappa_S$) where:
\begin{equation}
 F_{corr}=\frac{F_{obs}}{\kappa_S}
\end{equation}
Thus, all fluxes are normalized to a beam size of 32.5'' (i.e. 394 pc).  
In addition, we use the ground based $^{12}$CO, $^{13}$CO, HCN, HNC fluxes from \citet{1995A&A...302..343I} and Israel
(private communication), all normalized to a beam size of 32.5''.  We add the 27'' aperture integrated ground based observations of 
HCO$^+$ 1-0 and 4-3 transitions from \citet{2007ApJ...666..156K}, also normalized to a beam size of 32.5''.
Fluxes were first extracted using FTFitter (https://www.uleth.ca/phy/naylor/index.php?page=ftfitter), a program specifically created to extract line fluxes from SPIRE FTS spectra.  This is an IDL based GUI, which allows the user to fit lines, choose line profiles, fix any line parameter, and extract the flux.
We define a polynomial baseline to fit the continuum and derive the flux from the baseline subtracted spectrum.  In order to more accurately 
determine the amplitude of the line, we fix the FWHM for transitions higher than J=8-7 to the expected line width of $^{12}$CO at each source,
using the velocity widths measured by \citet{1995A&A...302..343I}.  

In the case of very narrow linewidths ($\sim780$ km/s at 650 $\mu$m), more narrow than the instrumental resolution (FWHM$_{max}$=1000 km/s at 650 $\mu$m), we do not fix
the FWHM but fit the lines as an unresolved profile, which is the case for CO J=4-3 through J=8-7.  We use an error of 30\% for our fluxes, which encompasses our dominant 
source of error, specifically the uncertainty of the beam size correction using the 450 $\mu$m SCUBA map (15\%) and line flux extraction (10\%). We also have some 
uncertainty in the SPIRE calibration error of $\sim$5\% for extended sources.

\section{Results}
\label{sec:res}

\begin{figure*}
\centering
\includegraphics[width=17cm]{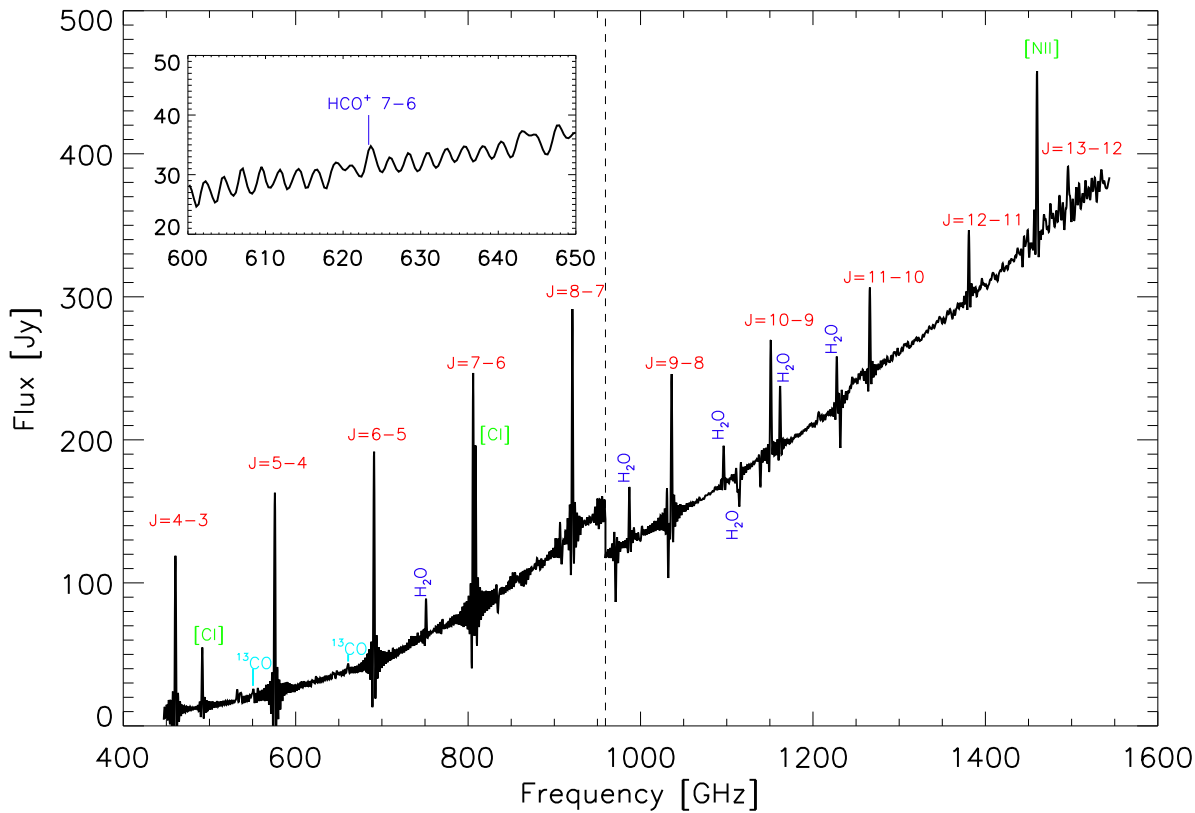}
\caption{SPIRE spectra of NGC 253, short wavelength and long wavelength bands separated by a black dotted line.  $^{12}$CO transitions are 
marked in red, $^{13}$CO transitions marked in light blue, atomic transitions are marked in green and H$_2$O and HCO$^+$ lines are marked in dark blue.}
\label{fig:spectra}
\end{figure*}

The spectra of NGC 253 are presented in Figure~\ref{fig:spectra}. The $^{12}$CO transitions are visible from J=4-3 to 
J=13-12 and labeled in red and the two $^{13}$CO transitions detected are marked in light blue.  There is also a strong 
detection of [NII] at 1461 GHz and [CI] at 492 GHz and 809 GHz in the rest frame, shown in green.  We detect 5 
strong water emission lines and H$_2$O $1_{11}-0_{00}$ in absorption, marked in blue.  We also detect HCO$^+$ 7-6, also marked in blue.  
In addition, we also find CH$^+$, CH, OH$^+$, H$_2$O$^+$, and HF, but do not point them out since they will not be used in our analysis.  As seen in 
Figure~\ref{fig:spectra}, there is a discontinuity between the high and low frequency bands of the spectrograph, 
marked with a dotted line.  This discontinuity is due to the fact that for the high frequency band, the beam size is much smaller than for the low frequency band.  
A beam correction factor ($\kappa_S$) for each wavelength is calculated using the method described in Section~\ref{sec:obs1}, 
and displayed in Table~\ref{tab:flux}.  Also, the baseline ripple seen in the spectrum, specifically in the inset of Figure~\ref{fig:spectra}, is due 
to the sinc profile of the strong CO transitions and does not represent noise.

\begin{table}
\caption{Observed (uncorrected) line fluxes and beam size correction factors ($\kappa_S$) normalized to 32.5''.  The errors on all derived fluxes are 30\%. }
 \begin{tabular}{|l||c|c|}
\hline
Line &  $\kappa_{S_{A}}$ & Line Flux \\
     &     32.5''        & [10$^{-16}$ W m$^{-2}$]\\
\hline
$^{12}$CO 1-0 \tablefootmark{a}& -- &0.3\\
$^{12}$CO 2-1 \tablefootmark{a}& -- &2.4\\
$^{12}$CO 3-2 \tablefootmark{a}& --&7.3\\
$^{12}$CO 4-3&1.13&12.8 \\
$^{12}$CO 5-4&1.00&17.1 \\
$^{12}$CO 6-5&0.94&17.2 \\
$^{12}$CO 7-6&1.04&18.2 \\
$^{12}$CO 8-7&1.07&17.9 \\
$^{12}$CO 9-8&0.67&12.2 \\
$^{12}$CO 10-9&0.64&9.3 \\
$^{12}$CO 11-10&0.62&7.7  \\
$^{12}$CO 12-11&0.60&5.5  \\
$^{12}$CO 13-12&0.60&3.9  \\
\hline
$^{13}$CO 1-0 \tablefootmark{a}&--&0.02 \\
$^{13}$CO 2-1 \tablefootmark{a}&--&0.2 \\
$^{13}$CO 3-2 \tablefootmark{a}&--&0.8 \\
$^{13}$CO 5-4&1.00&0.9 \\
$^{13}$CO 6-5&0.96&0.7 \\
\hline
$[$CI$]^3P_1-^3P_0$&1.09&4.5\\
$[$CI$]^3P_2-^3P_1$&1.04&11.4 \\
$[$NII$]^3P_1-^3P_0$&0.60&29.6\\
\hline
HCO$^+$ 1-0 \tablefootmark{b}&0.88&0.006 \\
HCO$^+$ 4-3 \tablefootmark{b}&0.88&0.2 \\
HCO$^+$ 7-6 &1.00 &0.5\\
\hline

HCN 1-0 \tablefootmark{c}&0.72&0.009 \\
HCN 3-2 \tablefootmark{c}&0.62&0.2 \\
HCN 4-3 \tablefootmark{c}&0.47 &0.1\\
\hline

HNC 1-0 \tablefootmark{c}&0.72&0.008 \\
HNC 3-2 \tablefootmark{c}&0.62&0.09 \\

\hline

H$_2$O 2$_{11}$-2$_{02}$&4.73&1.4 \\
H$_2$O 4$_{22}$-3$_{31}$&1.79&0.5 \\
H$_2$O 2$_{02}$-1$_{11}$&9.13&2.7 \\
H$_2$O 3$_{12}$-3$_{03}$&6.24&1.9 \\
H$_2$O 3$_{12}$-2$_{21}$&2.68&0.8 \\
H$_2$O 3$_{21}$-3$_{12}$&9.65&2.9 \\
H$_2$O 4$_{22}$-4$_{13}$&1.08&0.3 \\
H$_2$O 2$_{20}$-2$_{11}$&7.43&2.2 \\
\hline

 \end{tabular}
\tablefoot{ \\
 \tablefoottext{a}{From \citet{1995A&A...302..343I}.} \\
 \tablefoottext{b}{From \citet{2007ApJ...666..156K}.} \\
 \tablefoottext{c}{From Israel (private communication)} \\
 }
\label{tab:flux}
\end{table}

\section{Dissecting the CO Excitation Ladder}
\label{sec:coladder}
In order to analyze the excitation conditions in NGC 253, we can create a 'CO ladder' or spectral line energy distribution, 
which plots the flux of each CO transition as a function of the upper J number.  Since we have multiple transitions of 
both $^{12}$CO and $^{13}$CO, we can use both in our analysis.  Since we take a beam size of 32.5'' ($\sim 400$ parsec diameter), we 
get emission from multiple phases of the ISM in one spectrum and cannot spatially separate the distinct ISM environments.  
Thus, when analyzing the properties of each 
ISM phase, it is important to realize that these properties are average representative values of the dominant ISM environments that are responsible for 
the particular emission.

\subsection{LVG Analysis}
To better understand the ISM excitation in the center of NGC~253, we
start by modeling the ground-based fluxes, i.e. the first four
$^{12}$CO and first three $^{13}$CO line intensities and ratios with
the RADEX large velocity gradient (LVG) radiative transfer models \citep{2007A&A...468..627V}.  These
codes provide model line intensities as a function of three input
parameters per molecular gas phase: molecular gas kinetic temperature
$T_{\rm k}$, density $n$(H$_2$), and the CO velocity gradient $N({\rm
CO})$/d$V$.  By comparing model to observed line {\it ratios}, we
identify the physical parameters best describing the actual
conditions.

\nobreak In the modeling, we assume a constant CO isotopical
abundance $^{12}$CO/$^{13}$CO = 40 throughout. This is close to values
generally found in starburst galaxy centers
\citep{1991A&ARv...3...47H}.  We identify acceptable fits by exploring a large
grid of model parameter combinations ($T_{\rm k}$ = 10 - 150 K,
$n(_{H_2})$ = $10^{2}$ - $10^{5}$ cm$^{-2}$, and \textbf{$N$(CO)$/$d$V$ =
$6\times10^{15}$ - $3\times10^{18}$ cm$^{-2}$ / km s$^{-1}$)} for (combined) line ratios
matching those observed. No single-phase gas provides a good fit to
the observed line intensities. Consequently, we have also modeled the
$^{12}$CO and $^{13}$CO lines simultaneously with {\it two molecular gas
phases}, the relative contribution of the two phases being a free
parameter.  Lacking a detailed physical model for the distribution of
clouds and their sources of excitation, two phases is the most we can
fruitfully explore for J<5, especially because only the high signal-to-noise lower $J$
$^{13}$CO transitions have observations that allow us to break the
inherent $T_{k}-n$(H$_2$) degeneracy.  This degeneracy derives from the fact that the intensity ratios of optically
thick $^{12}$CO lines exhibit very little change
going from hot, low-density gas to cool, high-density gas (\emph{cf} Jansen 1995, Ph.D. Thesis Leiden University, Ch. 2).

No unique solution is obtained. However, in all combinations the
parameters of the first molecular gas phase are well-established and
narrowly constrained to a low kinetic temperature of about 60 K ($\pm 10$), and a
density of about log($n_{H}$) = 3.5 ($\pm 0.5$). The second phase has in all
combinations a reasonably well-determined higher density of typically
log ($n_{H}$) = 5.0, but the kinetic temperature is not at all
constrained by the lower $J$ transitions included, nor is the
proportion of molecular gas in either phase.

We now use this result, especially the firmly established parameters
of the coldest and least dense gas from phase 1, to optimize a 
\emph{three-phase} LVG model that includes the higher transitions as well:

\begin{equation}
\label{eq:mod}
Model=\Omega_ILVG_I+\Omega_{II}LVG_{II}+\Omega_{III}LVG_{III}
\end{equation}

\noindent where $LVG_I$, $LVG_{II}$, and $LVG_{III}$ are three LVG models of specific density, temperature, and column density in units of W m$^{-2}$.  $\Omega_I$, 
$\Omega_{II}$, and $\Omega_{III}$ represent the respective filling factors of each ISM phase.  Filling factors traditionally 
represent how much of the beam is filled, so they only range from 0 to 1.  However, in the case of our filling factors, we model one cloud with a 
$\delta v =1$ km s$^{-1}$ and allow for multiple clouds in our line of sight, such that $\Omega$ is not only a beam filling factor, but also a volume filling 
factor, which accounts for it being greater than one. In the case of $\Omega<1$, only a fraction of the beam is filled by clouds, whereas $\Omega > 1$ represents
the number of clouds in our beam volume, each with a slightly different velocity. We also attempted the same procedure for cloud models with 
$\delta v =5$ and 10 km s$^{-1}$ without any significant difference.  
We create a composite model for both $^{12}$CO and $^{13}$CO where each model ISM phase and filling factor is the same for $^{12}$CO and $^{13}$CO.  Since
we already have an idea of the first LVG phase (LVG I), we only vary the second and third LVG phases and all three filling factors.  

\begin{figure*}
\centering
\includegraphics[width=6cm]{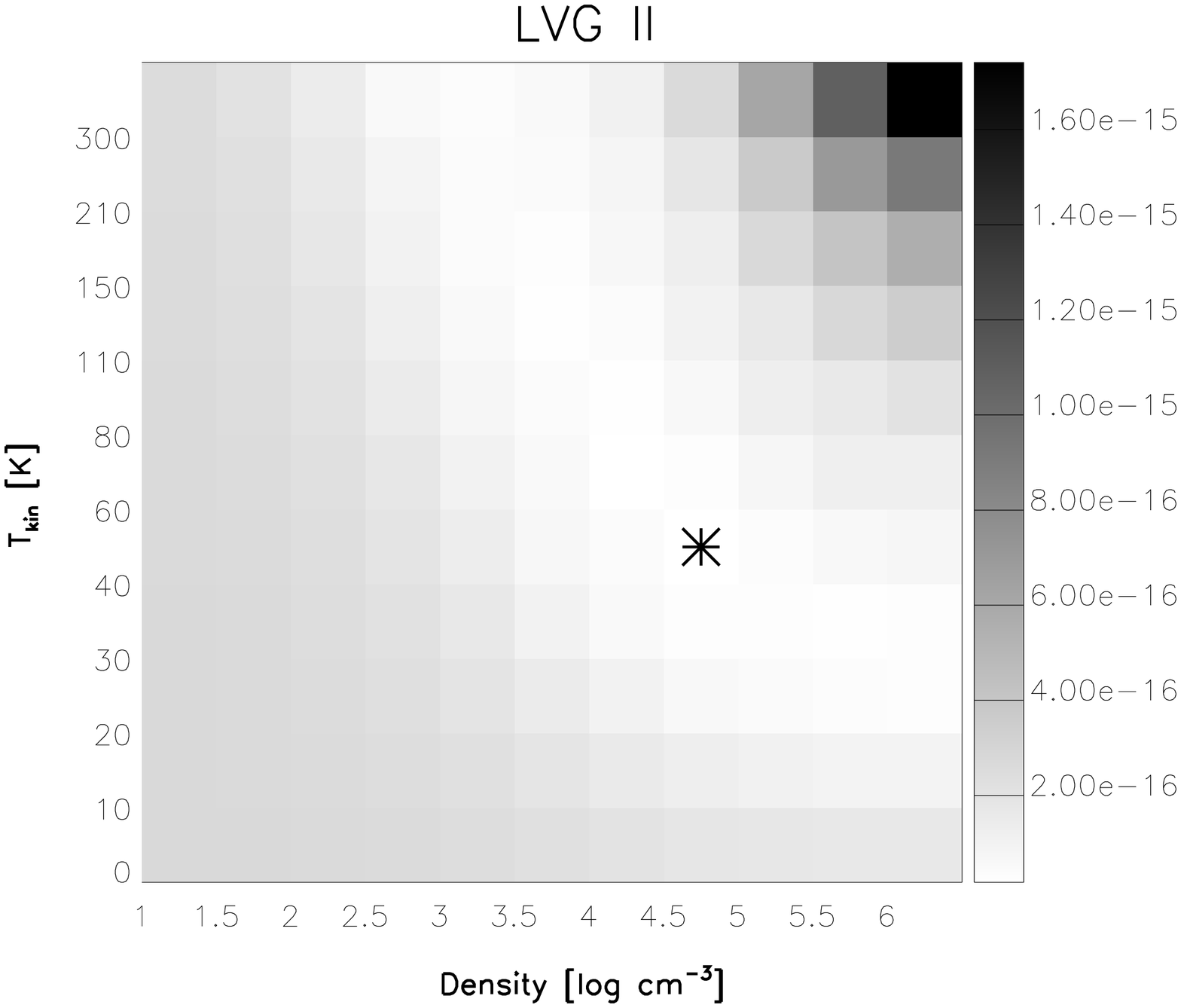}
\includegraphics[width=6cm]{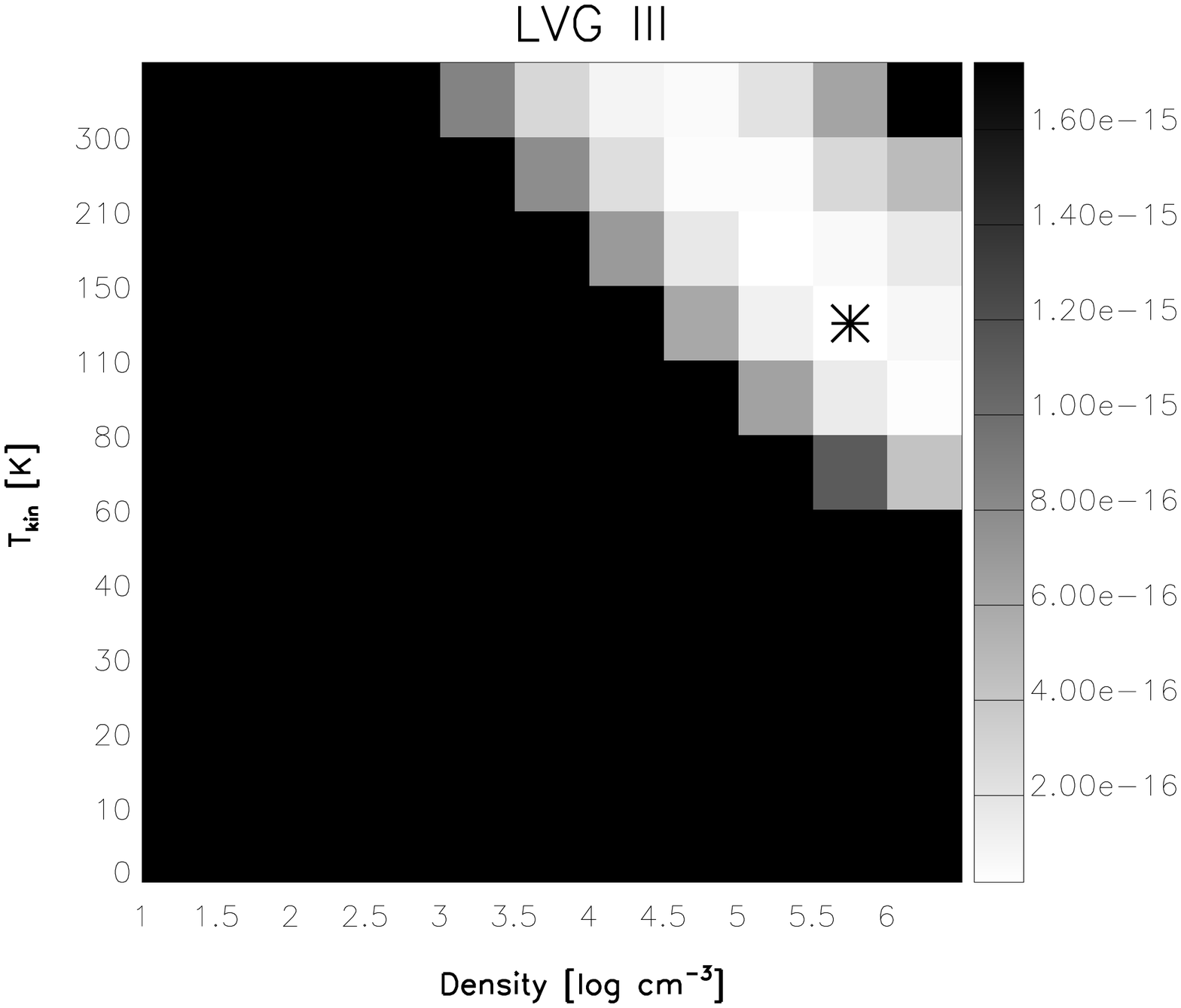}
\caption{Degeneracy plots of LVG II (right) and LVG III (left).  Each square represents a model with specific density 
(x axis) and temperature (y axis) for the best fitting $N($CO$)/\delta v$.  The color scale represents the $\chi^2$ value of that model fit to the CO data. \textbf{Since
we use a modified Pearson's $\chi^2$, the actual values of the grey scale are not meaningful, only their relative differences.}
The large black asterisk represents the best fitting model (minimum $\chi^2$). }
\label{fig:degens}
\end{figure*}

We perform a modified Pearson's $\chi^2$ minimized fit for $^{12}$CO and $^{13}$CO simultaneously, where the modified Pearson's $\chi^2$ is:
\begin{equation}
\chi^2=\frac{\sum_{i=1}^{N_{data}} (\frac{obs_i-model_i}{model_i})^2}{N_{data}}
\end{equation}
\noindent with \emph{obs$_i$} as the observed flux of a particular transition, \emph{model$_i$} is the composite model flux (Eq.~\ref{eq:mod}) 
of a particular transition, and \emph{N$_{data}$} is the total number of transitions.  Thus the $\chi^2$ represents a $\chi^2$ per transition for each molecule.  
We calculate the $\chi^2$ value as the sum of the $\chi^2$ for $^{12}$CO and $^{13}$CO for every possible combination of models in our grid.  
In doing this, we are able to plot the full parameter space for each phase, and show the $\chi^2$ value in gray scale for each combination of density and temperature
to see where the degeneracies lie.  Figure~\ref{fig:degens} shows these degeneracy plots for the second and third LVG phases (LVG II and LVG III).      

Although there is a best fit model, denoted by the black asterisk in Figure~\ref{fig:degens}, there are many models for each 
phase that have similarly low $\chi^2$ values, especially for LVG II.  In both LVG phases we see the general diagonal trend displaying the temperature-density degeneracy, trading 
temperature for density.  For LVG II, there is a large range of models that fit well, however the best fits tend to cluster at the low temperature, high 
density range (lower right corner).  On the other hand, LVG III fills a much smaller part of parameter space. The diagonal degeneracy still exists, but this phase 
requires both high temperature and high density to be fit well.  It is also important
to note that LVG II and LVG III occupy completely independent diagonal regions of parameter space, and since the diagonal trend trades temperature for density, 
we can think of these separate regions as unique pressure phases.  In Figure~\ref{fig:degens}, we only compare the degeneracies of temperature and density, yet
we also analyze the degeneracies of $N($CO$)/\delta v$ and find that within a factor of $\pm 5$ the column density is well constrained.
  
\begin{figure}
\resizebox{\hsize}{!}{\includegraphics{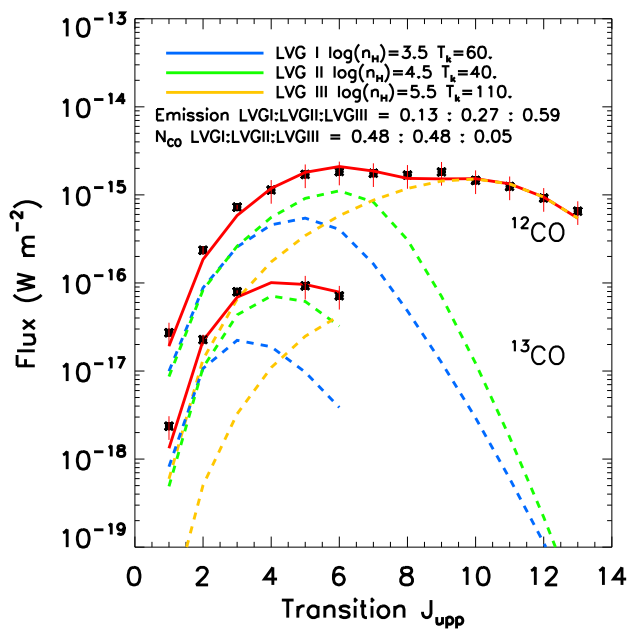}}
\caption{$^{12}$CO and $^{13}$CO excitation ladders of NGC 253 with flux of each transition plotted as black asterisks with red error bars.  
In blue, green and yellow dotted lines we plot the LVG I, II and III ISM phases with their filling factors.  The composite model is plotted 
with a red solid line.  The model density, temperature and column density are shown in the legend along with the relative contribution of 
each phase in terms of emission and column density.}
\label{fig:lvgfit}
\end{figure}

Although we cannot fully resolve the degeneracies, we will select a range of model parameters with a low $\chi^2$ value to continue our analysis.  We will define 
the density range 
of LVG II as $4<$log$_{10}(n_H/cm^{-3})<6$ and the range of LVG III as $4.5<$log$_{10}(n_H/cm^{-3})<6$.  Figure~\ref{fig:lvgfit} shows our three best fit ISM LVG phases.  
Here we do not include the dense gas tracers to constrain the density of our LVG models since the abundances may vary between ISM phases, however the densities of our best fit
LVG models are consistent with the HCN and HNC ratios shown in Figure~\ref{fig:co} and Table~\ref{tab:mcdr}.

The model parameters of our best three models (Figure~\ref{fig:lvgfit}) are summarized in Table~\ref{tab:lvg}.  The column density stated in this table is the model gradient,
log($N($CO$)/\delta v$). We show a relative contribution of the emission by summing the flux over all transitions of each model 
phase and comparing that to the flux of the composite model (red solid line).  We compare the total emission (summed flux of CO transitions) to the 
integrated emission of each ISM phase.  We find a relative contribution in terms of emission (luminosity) to the total CO flux from LVG I, LVG II, and LVG III
of $0.2:0.2:0.6$ respectively.  Next, we compare the relative scaled column densities which take the filling factors into account such that N$_i= \Omega_i \times N_{CO,i}$. 
We see a relative contribution in terms of column density of
$0.6:0.4:0.02$ for LVG I, II and III respectively, which is also proportional to the relative mass contributions of each phase. 
Although LVG III is responsible for the bulk of the line emission, and
all of the emission for the high J CO lines, it represents a very small fraction ($\sim2\%$) of the mass.  It is also interesting to note that LVG III contributes 
almost nothing to the $^{13}$CO emission.  This is due mainly to the low optical depth of LVG III, restricting the $^{12}$CO/$^{13}$CO to much lower than in LVG I and II. 
The fact that LVG III has such a high density, but such a low column density implies that the path length is very small or that the CO abundance is very low.  The low 
CO column density combined with a high temperature suggest the physical environment of a PDR where the high UV flux keeps most of the CO dissociated; we explore this 
further in Section 4.2.  We estimate a total gas mass of 3$\times10^7$ M$_\odot$, which is only a lower limit for the mass since we assume a [C]/[H] abundance
ratio of 1.4$\times10^{-4}$ and that all of the [C] is in CO.  However, in the case of a high-density, high-temperature LVG phase (LVG III), it is likely that the radiation 
environment is strong, and thus most of the CO would be dissociated, which would suggest a lower H$_2$ mass than there is in reality. Due to these uncertainties,
we will perform a more detailed
mass calculation in the following sections. 

\begin{table*}
\caption{Model parameters for the three LVG phases.}
 \begin{tabular}{|l||c|c|c|c|c|c|}
\hline
Component &  Density log(n$_{H}$) & Kinetic Temp & $log\frac{N_{CO}}{\delta v}$ & $\Omega$\tablefootmark{a} & C$_{em}$\tablefootmark{b} & C$_{N_{CO}}$\tablefootmark{c}\\
          &  log[cm$^{-3}$]  & K   & log[cm$^{-2}$ / km s${^-1}$] & && \\
\hline
LVG I   & 3.5 & 60  & 17 & 5.0 & 0.13 & 0.48 \\
LVG II  & 4.5 & 40 & 17 & 5.0 & 0.27 & 0.48  \\
LVG III & 5.5 & 110 & 17 & 0.5 & 0.59 & 0.05 \\
\hline
 \end{tabular}
 \tablefoot{ \\
 \tablefoottext{a}{$\Omega$ is the beam filling factor for each ISM phase.} \\
 \tablefoottext{b}{C$_{em}$ is the fractional contribution of each ISM phase to the emission.} \\
 \tablefoottext{c}{C$_{N_{CO}}$ is the fractional contribution of each ISM phase to the column density.} \\
 }
\label{tab:lvg}
\end{table*}

\subsection{PDR Analysis}
\label{sec:pdrs}
Now that we have constrained the density and temperature ranges of the three dominant molecular ISM phases, we can determine 
which physical processes are exciting the molecular gas. Since the nucleus of NGC 253 is experiencing high star formation rates,
there is an increase in number density of OB stars and thus a high UV energy density.  Through photoelectric heating and FUV pumping of H$_2$, the 
FUV photons heat the outer layers (A$_V$<5) of molecular clouds.  This area of the molecular cloud is the PDR, and is responsible
for warm molecular gas emission.  The thermal state of PDRs is determined by processes such as photo-electric heating, heating by pumping of H$_2$, followed by collisional de-excitation, 
heating by cosmic rays, [OI] and [CII] fine-structure line cooling, and CO, H$_2$O, H$_2$, and OH molecular cooling. The ionization degree of the gas is driven by FUV
photo-ionization, and counteracted by recombination and charge transfer reactions with metals and PAHs. The ionization degree is at most $x_{\rm e} \sim 10^{-4}$ outside of the fully ionized zone. 
The chemistry exhibits two fundamental transition, H to H$_2$, and C$^+$ to C to CO.  
Using PDR models from \citet{2012A&A...542A..65K}, which solve for chemistry and thermal balance throughout
the layers of the PDR, we use the predictions of the $^{12}$CO and $^{13}$CO emission as a function
of density, radiation environment ($G$, the Habing radiation field), and column density.  The density profile is constant, as in the LVG models, and the Habing field 
is defined as 1 $G$=G$_0$=1.6$\times 10^{-3}$ 
erg cm$^{-2}$ s$^{-1}$ from photons between 6 and 13.6 eV. The fact that these models use $G$ instead
of temperature provides a 
direct diagnostic of the radiation source responsible for the gas excitation. The model is a semi-infinite slab model which is evaluated at A$_V$=3,4,5,10,20, and 30 mags, 
and we calculate the column density from the A$_V$ and density that gives us the best fit emission. 

We narrow the PDR model parameter space by allowing only the densities based on Figure~\ref{fig:degens} ($4<$log$_{10}(n_H/cm^{-3})<6$) for the second phase 
and ($4.5<$log$_{10}(n_H/cm^{-3})<6$) for the third phase.  In order to further constrain our fits, we include
$^{12}$CO, $^{13}$CO, HCN, and HNC to our $\chi^2$ calculation.  Since HCN and HNC are typically dense gas tracers (both have n$_{crit}>10^5$ cm$^{-3}$), they provide an important 
additional constraint on the second
and third phases, since according to our LVG analysis both phases are high density.  We also have multiple HCO$^+$ transition observations but we do not use them
in our analysis because of the complicated nature of the emission of HCO$^+$.  Its abundances are very sensitive to the interplay between heating and ionization rate and the
resulting chemical state of the gas \citep{2011A&A...525A.119M} and thus an untrustworthy diagnostic.  

\begin{figure*}
\centering
\includegraphics[width=17cm]{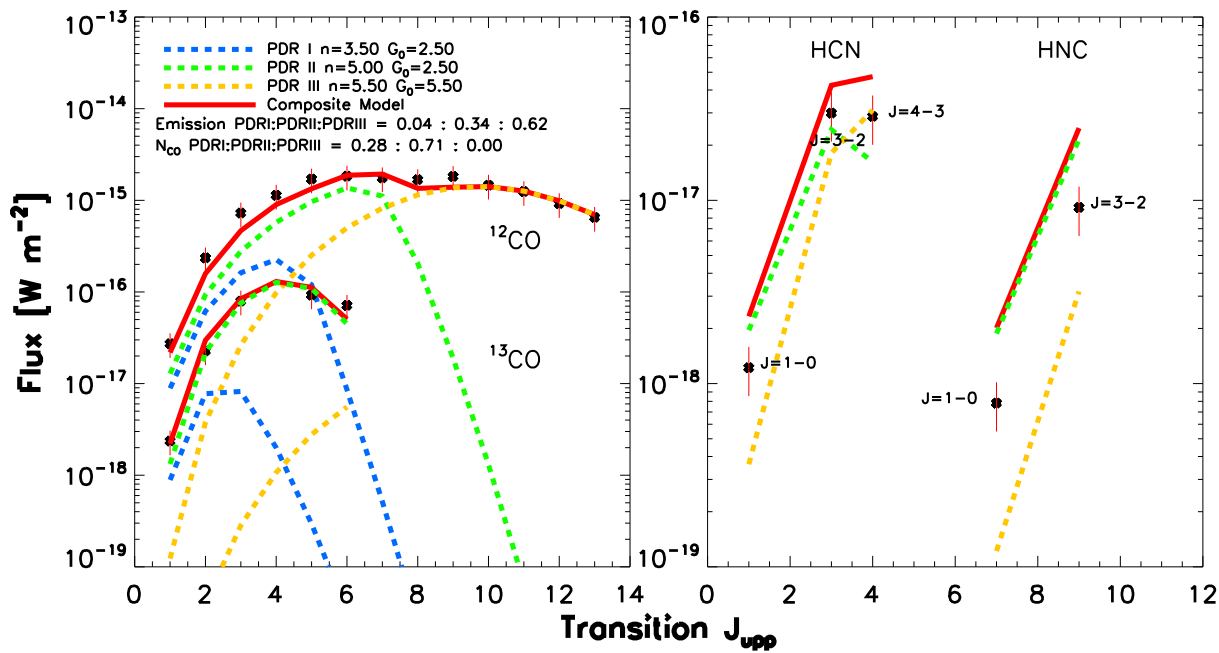}
\caption{Left: $^{12}$CO and $^{13}$CO excitation ladders, right: HCN and HNC excitation ladders of NGC 253 with flux of each transition plotted as black asterisks 
with red error bars.  In blue, green and yellow dotted lines we plot the PDR I, II and III ISM phases with their filling factors.  The composite model is plotted 
with a red solid line.  The model density, G$_0$ and column density are shown in the legend along with the relative contribution of 
each phase in terms of emission and column density.}
\label{fig:pdr}
\end{figure*}

The parameters of the best fitting models are listed in Table~\ref{tab:pdr}.  In Figure~\ref{fig:pdr} the PDR fit is shown for
$^{12}$CO, $^{13}$CO, HCN, and HNC.  Each PDR phase is represented by a blue, green
, and yellow dotted line, while the composite model is shown as a red solid line.  We can achieve a relatively good fit for all phases, yet
it is important to check if this fit makes physical sense.   
     
\begin{table*}
\caption{Model parameters for the three PDR phases.}
 \begin{tabular}{|l||c|c|c|c|c|c|c|c|}
\hline
Component &  Density log(n$_{H}$) & log(G$_0$) & log(N$_{CO}$) & log(N$_{CO}$) & $\Omega$\tablefootmark{a} & C$_{em}$\tablefootmark{b} & C$_{N_{CO}}$\tablefootmark{c} & Mass$_{N_{H_2}}$\tablefootmark{d} \\
          &  log[cm$^{-3}$]  &    & log[cm$^{-2}$] & log[cm$^{-2}$]& & &M$_\odot$& M$_\odot$ \\
\hline
PDR I   & 3.5 & 2.5 & 17.1 & 21.5 & 10.0 & 0.04 & 0.28 & $9\times10^7$\\
PDR II  & 5.0 & 2.5 & 17.5 & 21.5 & 10.0 & 0.34 & 0.71 & $9\times10^7$\\
PDR III & 5.5 & 5.5 & 16.0 & 21.2 & 1.5 & 0.62 & $<0.01$ &$7\times10^6$\\
\hline
 \end{tabular}
  \tablefoot{ \\
 \tablefoottext{a}{$\Omega$ is the beam filling factor for each ISM phase.} \\
 \tablefoottext{b}{C$_{em}$ is the fractional contribution of each ISM phase to the emission.} \\
 \tablefoottext{c}{C$_{N_{CO}}$ is the fractional contribution of each ISM phase to the CO column density.} \\
 \tablefoottext{d}{Mass$_{N_{H_2}}$ is the mass of each ISM phase as estimated by the column density using Eq.~\ref{eq:col}.} \\
 }
\label{tab:pdr}
\end{table*}

Using these three model PDR phases, we can estimate the molecular gas mass in each phase by:
\begin{equation}
\label{eq:col}
M_{H_2}=\sum_{i}^{n}\frac{\Omega_i N_{H_2,i}A_{beam}m_{H_2}}{M_\odot}
\end{equation}  

\noindent where $N_{H_2}$ is the H$_2$ column density in cm$^{-2}$ which is consistently calculated in the PDR models, $A_{beam}$ is the 
beam area in cm$^2$, and $m_{H_2}$ is the mass of a hydrogen molecule.
The mass of each PDR phase is shown in Table~\ref{tab:pdr} and the total mass of the system is $1.9
\times10^8$ M$_\odot$, which is almost equal to the gas mass measured for a beam size of 80'', $3\times10^8$ M$_\odot$ \citep{2008A&A...490...77W}.
\citet{1999MNRAS.303..157H} and \citet{2003ApJ...586..891B} found a gas mass of $\sim3\times10^7$ M$_\odot$ for a beam size of 15'', roughly half the size of our 
beam yet we find a gas mass 6.5 times larger.  In addition, our LVG models predict a mass of $\sim3\times10^{7}$ M$_\odot$, which is a reasonable (albeit a lower limit) gas mass 
estimate.  The LVG mass is a lower limit since we can see for a PDR with a high radiation field, the CO becomes dissociated.  Thus using the standard abundance 
of CO to H$_2$ will under estimate the true H$_2$ column. 

To further test the physicality of our PDR models by comparing the atomic gas to the molecular gas.  \citet{1994ApJ...423..223C} uses 
[CII] 158 $\mu$m to trace the atomic gas mass in the inner 40'' of NGC 253 and finds an atomic mass of $2.4\times10^6$ M$_\odot$.  This atomic mass is 
30 times smaller than the molecular gas that we observe in our 32.5'' beam, which is similar to the findings of \citet{2003ApJ...586..891B} and \citet{2008ApJ...689L.109H}.
\citet{1985ApJ...291..722T} predict that 
in Galactic PDRs, the first $\sim$ 3 A$_V$s are irradiated and then cooled via atomic lines, whereas the next 3 A$_V$ are 
cooled through molecular lines.  From this they estimate that in a PDR, the masses of atomic and molecular mass should be about even.  Since our molecular mass is 30 times
greater than our atomic mass, we can calculate the A$_V$ where the atomic heating must transition to molecular heating to preserve this ratio and in which conditions
this takes place.  

First, we assume our PDR is a 2-D finite slab with an (ionized) atomic outer layer and a molecular deeper layer.  UV radiation is penetrating the slab from one side 
and the slab has a constant density and a finite depth.  In the case of our PDRs, the depth is A$_V$=5, as shown in Table~\ref{tab:pdr}.  We can measure 
the mass in each layer, which is proportional to Volume$\times$Density, and we assume a constant density and beam area (V=BA$\times $r), where r is the depth into the cloud, so these both drop out.  
The volume of the first layer is proportional to $R_x$, where R$_x$ is the radius where the 
atomic to molecular transition occurs.  The volume of the molecular layer is proportional to $R_{A_V=5}-R_x$, where R$_{A_V=5}$ is the distance to the end of the slab. 
We can than say that the ratio of the molecular to atomic mass is proportional to $(R_{A_V=5}/R_x)-1$ and we observe this ratio to be 30.  Solving this equation for 
R$_x$, we find that in order to observe a molecular to atomic mass ratio of 30, the transition between atomic to molecular must occur at A$_V$=0.2.  Although this is a 
simplified calculation, it demonstrates the difficulty in producing 30 times more molecular mass than atomic.  We use our PDR models to test in which environments this 
very low transition depth occurs.  We find the lowest transition depth in our model parameter space is for a PDR with the highest density (10$^6$ cm$^{-3}$) and the lowest
radiation field (10$^2$ G$_0$).  The transition depth for our three PDR models is A$_V$= 3.7, 2.2, and 4.5 for PDR I, II, and III respectively.  Since all of these are 
much above the necessary A$_V$=0.2, we suggest that an additional heating source is necessary in order to account for the large amount of bright molecular gas.

In order to determine the cause of this discrepancy, we start with our most constrained ISM phase, LVG/PDR I.  We have determined that this ISM phase has a density 
of 10$^{3.5}$ cm$^{-3}$ and a kinetic temperature of 60 K.  We fit this with a PDR with the same density, same column density, and a $G=10^{2.5}$ G$_0$.  From G$_0$ 
we can calculate the average temperature of the PDR, which is the dominant factor in determining the emission intensity of the lines.  We find that in PDR I, 
we only have a temperature of 13.5 K, while we need a temperature of 60 K.  If we increase the radiation field in order to heat the gas further, we get up to a peak 
temperature of 18 K and then begin to dissociate the CO.  The only way in which we can reproduce a PDR with n=10$^{3.5}$ cm$^{-3}$ and T=60 K is by including an 
additional heating term.  
 
\section{Dominant Molecular Excitation Mechanisms}
\label{sec:highj}
Since our analysis suggests that a PDR can not be the sole source of the high J CO transitions, we investigate
excitation from other sources for comparison.
\subsection{Excitation Mechanisms}
Aside from PDR heating, the three dominant heating mechanism of molecular gas are X-rays, cosmic rays, and  mechanical heating. 
X-rays heat gas in regions (called X-ray Dominated Regions, XDRs) similar to PDRs except that the chemistry is driven by X-ray photons
instead of FUV photons, which are able to penetrate further into the cloud without efficiently heating the dust at the same time.  These X-rays are mostly produced by
active galactic nuclei (AGN) or in areas of extreme massive star formation.  The observed X-ray flux between 0.1 and 2.4 keV is 5.72$\pm$0.59$\times 10^{-12}$ erg
s$^{-1}$ cm$^{-2}$, which
translates to a luminosity of L$_X$ = 4$\times 10^{39}$ ergs s$^{-1}$ at a distance of 2.5 Mpc \citep{2000A&A...360...24P}. Since NGC 253 lacks an X-ray bright AGN and 
has a relatively low X-ray luminosity compared to regions where we see X-ray heating, 
we exclude this mechanism from further analysis.  
Cosmic rays can also heat gas in Cosmic Ray Dominated Regions (CDRs).  Cosmic rays are able to penetrate into the very centers of molecular clouds, where 
 even X-rays have trouble reaching and are typically produced by supernovae.  Similarly, PDRs with additional mechanical heating (mPDRs) are due to turbulence in the ISM, 
 which may be driven by supernovae, strong stellar winds, or jets.


In order to test for mechanical heating, we can add a mechanical heating term to the PDR models (mPDR models).  
Since turbulence can penetrate a cloud at all scales 
and depths, as a simple approximation the mechanical heating ($\Gamma_{mech}$) is introduced uniformly throughout the 1-D PDR models.
We parameterize the strength of $\Gamma_{mech}$ with $\alpha$, which represents the fractional
contribution of mechanical heating in comparison to the total heating at the surface of a pure PDR (excluding mechanical heating).  At the surface the heating budget is dominated by
photoelectric heating.  When $\alpha$=0, there is only photoelectric heating and cosmic ray heating and when $\alpha$=1, the mechanical heating
rate is equal to the photoelectric heating at the edge of the cloud.  These rates can be related
to supernovae rates, which in turn can be related to star formation rates
(as was done in \citet{2008A&A...488L...5L}). The freedom in the choice of
the values of $\alpha$ allow for a flexibility in modeling the source
of the absorbed mechanical heating. The surface heating rates for the
PDR models can be recovered from Figure 1 in \citet{2012A&A...542A..65K}.
More details about relating the mechanical heating to the star formation rate can
be found in the methods section of that paper and in \citet{2008A&A...488L...5L}.  In addition to the mechanical heating, there is a heating rate due to 
cosmic rays applied throughout the PDR.  The heating by cosmic rays is only dependent on H$_2$ density and ionization rate, in this case the galactic 
cosmic ray rate (0.2-3.5 $\times 10^{-16}$ s$^{-1}$) \citep{1978ApJ...222..881G,1986ApJS...62..109V,2012ApJ...745...91I} and the H$_2$ density of the selected model.

Since both cosmic rays and mechanical heating derive from supernovae, it is likely that if we have excitation from one mechanism, we should expect excitation from both.  Thus, in
addition to the mPDR models, we have run mCDR models.  These models not only have photoelectric heating from the PDR and mechanical heating, they also have an increased 
cosmic ray ionization rate of 750 $\zeta_{gal}$ or 3.75$\times10^{-14}$ s$^{-1}$, 
as suggested by \citet{2003ApJ...586..891B}. The enhanced cosmic ray ionization rate is paired with a varying mechanical heating rate from 
$\alpha=0$, or no mechanical heating, up to $\alpha=1.0$ which translates to a heating rate of $\Gamma_{mech}$=1.3$\times10^{-15}$ erg s$^{-1}$ cm$^{-3}$.  
 This allows for pure cosmic ray
 excitation, where $\alpha=0$ as well as the case where both mechanical heating and cosmic rays play a role in the gas heating.

\subsection{Application to NGC 253}
\begin{figure}
\resizebox{\hsize}{!}{\includegraphics{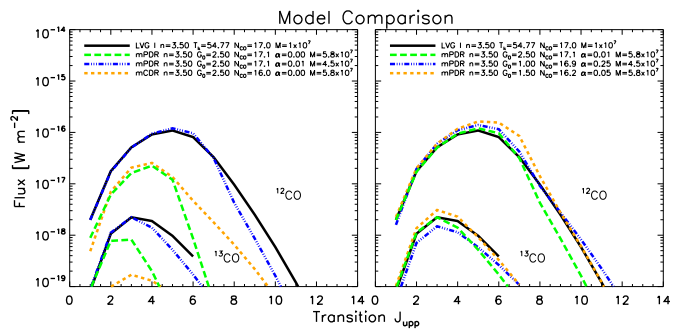}}
\caption{Comparison of PDR (green), mPDR (blue), and mCDR (orange) models to LVG I (black) for both $^{12}$CO and $^{13}$CO.}
\label{fig:comp}
\end{figure}

In order to determine the alternate heating mechanism, we first start with our most constrained ISM phase, PDR I/LVG I.  With only photoelectric heating, we cannot 
reach the temperature measured in the LVG model (60 K), and thus must include an additional heating source.  In order to compare various heating sources, we 
plot the LVG model (Figure~\ref{fig:comp}, left panel) in black and overplot other similar models.  Similarly to the pure PDR models, pure CDR with no mechanical heating dissociates the CO
before the temperature can get high enough to produce the observed emission.
First, we overplot the PDR model that we used in the PDR fit (Section 4.2) in green.  The emission from the pure PDR model is approximately a factor of 10 
lower than the LVG model of the same density and T=60 K.  This explains the high filling factors necessary to fit the model emission to our observations, and thus the 
high mass estimates.  We can then use the same parameters as the pure PDR fit, but add additional heating to see if we can reproduce the LVG I emission profile.  First we try 
to fit the emission with an CDR, identical to the PDR except with an enhanced cosmic ray rate of 750 times the galactic CR rate. This curve is overplotted in orange. Not only is the $^{12}$CO
10 times less luminous, the $^{12}$CO/$^{13}$CO ratio is not preserved, the $^{13}$CO is underproduced in comparison to the LVG I 
model, which makes this an unlikely model.  We can then test a mechanically heated PDR (mPDR) with an $\alpha$ of only 1\%.  This matches the emission of LVG I for both $^{12}$CO and $^{13}$CO 
and requires only a small amount of mechanical heating.  Therefore, we choose mechanical heating as the necessary additional heating mechanism for the first ISM phase.

Next, we must test this first component to see if it is well constrained.  In the right panel of Figure~\ref{fig:comp}, we have varied radiation strength, column density, and 
amount of mechanical heating to show that this component is not well constrained.  We can match the emission of the LVG model with any model with mechanical heating between 1-25\% and radiation 
between 10$^{1.0-2.5}$ G$_0$.  However, it is important to note that while these parameters change, so does the column density in a way that preserves mass. Therefore, we cannot determine 
which of these models represents the true physical conditions, but we can say that mechanical heating is necessary and the mass of the component is around
5$\times10^7$ M$_\odot$.  In order to proceed with the analysis, we select the model with 1\% mechanical heating (blue in Figure~\ref{fig:comp}) since it is a slightly better fit to the $^{13}$CO. 

\begin{figure*}
\centering
\includegraphics[width=17cm]{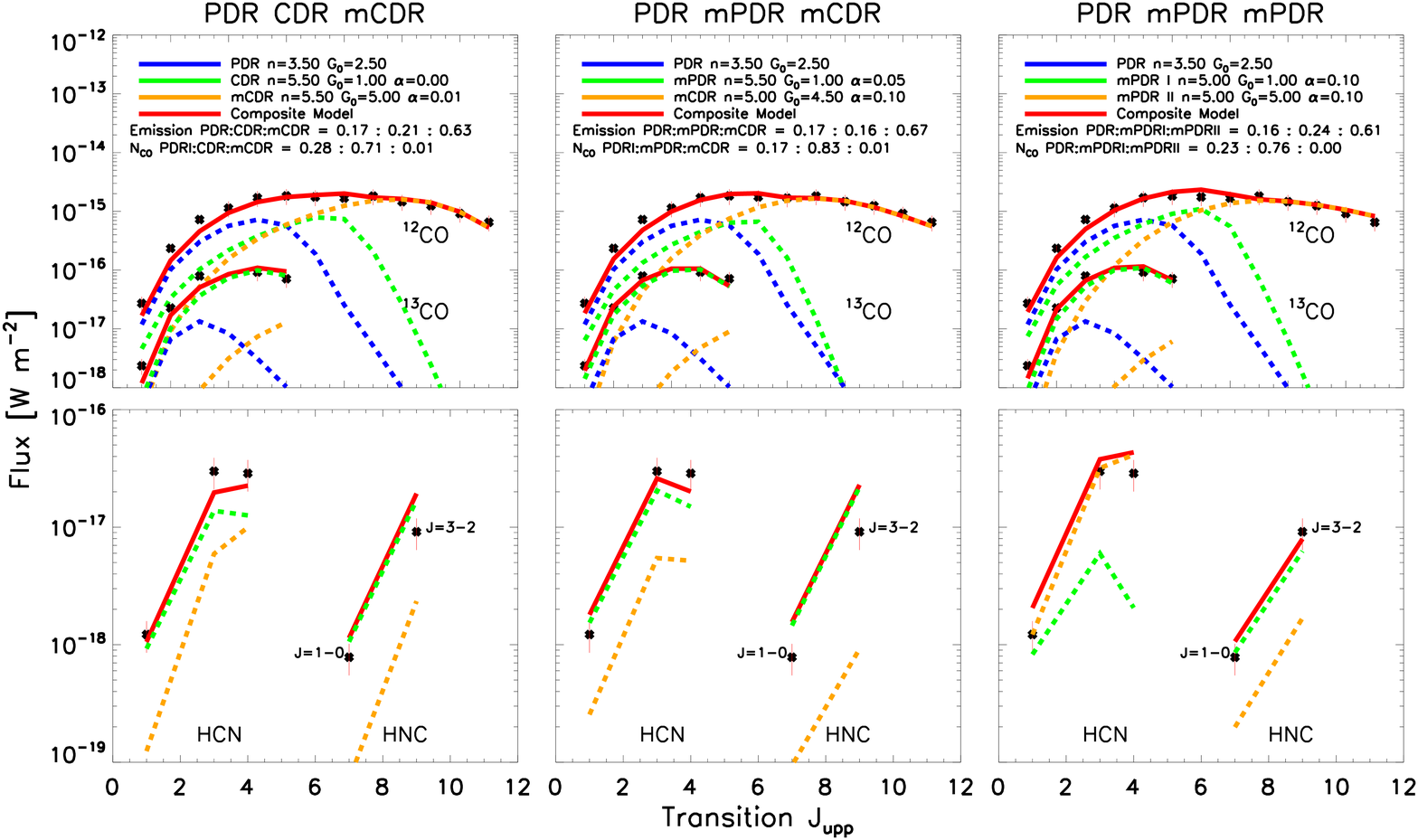}
\caption{$^{12}$CO and $^{13}$CO (top) with HCN and HNC (bottom) excitation ladders of NGC 253 with flux of each transition plotted as black asterisks with red error bars.  
In blue, green and yellow dotted lines we plot the PDR/mCDR/mCDR (left), PDR/mPDR/mCDR (center), and PDR/mPDR/mPDR(right) ISM phases with their filling factors.  
The composite model is plotted with a red solid line.  The model H$_2$ density [log cm$^{-3}$], G$_0$ [log Habing flux], and
percentage mechanical heating ($\alpha$) are shown in the legend along with the relative contribution of 
each phase in terms of emission and column density.}
\label{fig:co}
\end{figure*}

Now that we have determined the range of models applicable for the first ISM phase, we fit our second and third ISM phases with the mPDR and mCDR models to determine their dominant heating
source.  We hold the first ISM phase constant and equal to that of the 1\% mechanical heating mPDR in Figure~\ref{fig:comp} but we allow the 
filling factor to vary.  The second and third phases are held to the same requirements as in the PDR fitting, namely 
the second phase density is in between 10$^4$ and 10$^6$ cm$^{-3}$ and the third phase density is in between 10$^{4.5}$ and 10$^6$ cm$^{-3}$.
The $\chi^2$ for the $^{12}$CO, $^{13}$CO, HCN, and HNC ladders are summed to find an all around best fit.  The best
fitting models are shown in Figure~\ref{fig:co} for:
\begin{enumerate}
 \item 1 mPDR and 2 mCDRs
 \item 2 mPDRs and 1 mCDR
 \item 3 mPDRs.
\end{enumerate}
The parameters of the fit for each case are shown in Table~\ref{tab:mcdr}.

\begin{table*}
\caption{Model parameters for the three ISM phases for each of the three cases.}
 \begin{tabular}{|l||c|c|c|c|c|c|c|c|c|}
\hline
Component &  Density log(n$_{H}$) & log($G$) & log(N$_{CO}$) &log(N$_{H_2}$)& $\alpha$ & $\Omega$\tablefootmark{a} & C$_{em}$\tablefootmark{b} & C$_{N_{CO}}$\tablefootmark{c} & Mass$_{N_{H_2}}$\tablefootmark{d} \\
          &  log[cm$^{-3}$]  &  G$_0$  & log[cm$^{-2}$] &log[cm$^{-2}$]& \%& && &M$_\odot$ \\
\hline
\multicolumn{9}{|c|}{\textbf{Case 1}      M$_{tot}$: $8.4\times10^7$ M$_\odot$ } \\
\hline
mPDR     & 3.5 & 2.5 & 17.1 & 21.5&1 &6.0 & 0.17 & 0.28 & $5.7\times10^7$\\
CDR   & 5.5 & 1.0 & 17.8 &21.4 &0 &3.0 & 0.21 & 0.71 & $2.3\times10^7$\\
mCDR  & 5.5 & 5.0 & 16.8 &21.4 &1 &0.5 & 0.63 &0.01 &$3.8\times10^6$\\
\hline
\multicolumn{9}{|c|}{\textbf{Case 2}      M$_{tot}$: $1.1\times10^8$ M$_\odot$} \\
\hline
mPDR I  & 3.5 & 2.5 & 17.1 & 21.5&1 &6.0 & 0.17 & 0.17 & $5.7\times10^7$\\
mPDR II & 5.5 & 1.0 & 17.8 &21.4 &5&5.0 & 0.16 & 0.83 & $4.5\times10^7$\\
mCDR & 5.0 & 4.5 & 16.2 & 21.2&10 &1.5 & 0.67 & $<0.01$ &$7.1\times10^6$\\
\hline
\multicolumn{9}{|c|}{\textbf{Case 3}      M$_{tot}$: $1.1\times10^8$ M$_\odot$} \\
\hline
mPDR I   & 3.5 & 2.5 & 17.1 & 21.5&1 &6.0 & 0.16 & 0.23 & $5.7\times10^7$\\
mPDR II  & 5.0 & 1.0 & 17.7 & 21.4&10&5.0 & 0.24 & 0.76 & $3.8\times10^7$\\
mPDR III & 5.0 & 5.0 & 15.5 &21.0 &10&5.0 & 0.61 & $<0.01$ &$1.5\times10^7$\\
\hline
 \end{tabular}
  \tablefoot{ \\
 \tablefoottext{a}{$\Omega$ is the beam filling factor for each ISM phase.} \\
 \tablefoottext{b}{C$_{em}$ is the fractional contribution of each ISM phase to the emission.} \\
 \tablefoottext{c}{C$_{N_{CO}}$ is the fractional contribution of each ISM phase to the column density.} \\
 \tablefoottext{d}{Mass$_{N_{H_2}}$ is the mass of each ISM phase as estimated by the column density using Eq.~\ref{eq:col}.} \\

 }
\label{tab:mcdr}
\end{table*}

\citet{2003ApJ...586..891B} estimate an 
ionization rate of (1.5-5.3)$\times10^{-14}$ s$^{-1}$, which is 750 times larger than the galactic ionization rate.  \citet{2009Sci...326.1080A} measured a cosmic ray ionization rate up to 
1000 times our galactic ionization rate, which is consistent with the results from \citet{2003ApJ...586..891B}.  We consider both of these cosmic ray ionization rates to be upper limits, and suggest that the true cosmic ray ionization rate is somewhere in between galactic and 750 times the galactic value.  The difference between a cosmic ray ionization rate of 1000 and 750 times the galactic value is negligible, so we will use the values 750 throughout the rest of the paper.
\citet{2006ApJS..164..450M} finds evidence that large scale, low velocity shocks drive the chemistry and heating of NGC 253.  In support of this, 
diffuse SiO emission has been detected throughout the nuclear region \citep{2000A&A...355..499G}.  
In addition, \citet{2005ApJ...620..210M} have measured an over-abundance of OCS, which is caused by low velocity shocks injecting these molecules into the gas phase. 
Mechanical heating can also be introduced into the ISM through supernova remnants \citep{2008A&A...488L...5L}
or strong jets and we test its efficiency at heating molecular gas using the mPDR and mCDR models.  Strong molecular outflows have been observed by \citet{2013Natur.499..450B}, and 
they estimate an outflow rate of 9 M$_\odot$ yr$^{-1}$, implying the outflows are a strong source of turbulence in the galactic nucleus. 

Since \citet{2003ApJ...586..891B} suggest that cosmic rays are responsible for most of the molecular gas 
excitation in NGC 253, we first use case 1, two mCDR models, to fit our observations and see if they are consistent with cosmic ray excitation. 
We define cosmic ray excitation as
ionization by cosmic rays, which produces slow electrons that excite atomic and molecular hydrogen, and in turn produces photons that heat the gas \citep{2012ApJ...756..157G}. As seen in the left panel of 
Figure~\ref{fig:co}, the best fitting model requires no mechanical heating for the second ISM phase, thus it is a pure CDR and requires only 1\% mechanical heating for the third ISM 
phase. The fit of case 1 to CO, HCN, and HNC is very good, yet this begs the question what if the first component also used an enhanced CR rate.  We tested a fit with all three ISM 
phases including an enhanced cosmic ray rate and similarly to Figure~\ref{fig:comp} the $^{13}$CO is not fit at all.  If we tweak the $\chi^2$ calculation to weight CO more important than 
HCN and HNC, we can fit $^{12}$CO and $^{13}$CO but completely destroy the fits of HCN and HNC. However, the fact that we cannot fit three components with a high CR rate does not mean that
the fit of case 1 is not possible.  Instead, it could represent the physical situation where the diffuse ISM, traced by the 
first ISM phase, has a diluted cosmic ray rate that can be assumed to be equivalent to the galactic rate.

Since cosmic rays are predominately from supernovae, they may only affect a small percentage of the total gas mass.  Therefore, we attempt to recreate our observed line fluxes 
in case 2 with one PDR, one mPDR, and one mCDR.  This scenario represents the case where mechanical heating is present throughout the ISM, and in addition, cosmic rays excite a small 
amount of gas around supernova remnants. The best fit models for this scenario are shown in the center panel of Figure~\ref{fig:co}.  The models fit the $^{12}$CO, 
$^{13}$CO, HCN, and HNC reasonably well, if not within the error bars of all transitions.  Similarly to case 1, the HNC emission is off by a factor of $\sim2$, yet since our 
models are not very advanced in their 
treatment of the chemistry, a factor of 2 is not enough to rule out this case.  

Finally, we test the case where mechanical heating is the dominant heating mechanism and cosmic rays are not important to the heating
of the molecular gas.  The right panel of Figure~\ref{fig:co} shows the best fit for case 3, three mPDR models, mPDR II has an $\alpha$=0.1 which translates into 
a $\Gamma_{mech}=1 \times 10^{-20}$ erg s$^{-1}$ cm$^{-3}$ and mPDR III has an $\alpha$=0.1 which is $\Gamma_{mech}=1.6 \times 10^{-18}$ erg s$^{-1}$ cm$^{-3}$.  
This model fits all the $^{12}$CO and $^{13}$CO transitions within the error bars.  In addition, HCN and HNC are better fit than in both case 1 and case 2.
Since all three cases reproduce very similar, and very good fits of the molecular transititons, we cannot differentiate between the precise mechanism.  
Still even in case 1 mechanical heating is needed not only in the first ISM phase but in the third ISM phase
needing an $\alpha$=0.01, which is a non-negligible heat rate (1$\times 10^{-18}$ erg s$^{-1}$ cm$^{-3}$) as shown in Figure~\ref{fig:heating}.  Thus it is clear that mechanical heating is 
necessary to reproduce the observed emission, regardless of the added heating of cosmic rays.  This finding is similar to that of \citet{2011ApJ...742...88N} in NGC 
891 and \citet{2008ApJ...689L.109H} for NGC 253 itself, where they use a combination of shock and PDR models to determine that microturbulence and shocks are responsible for heating the mid-J CO gas. 

\section{Discussion}
\label{sec:disc}
\begin{figure*}
\centering
\includegraphics[width=17cm]{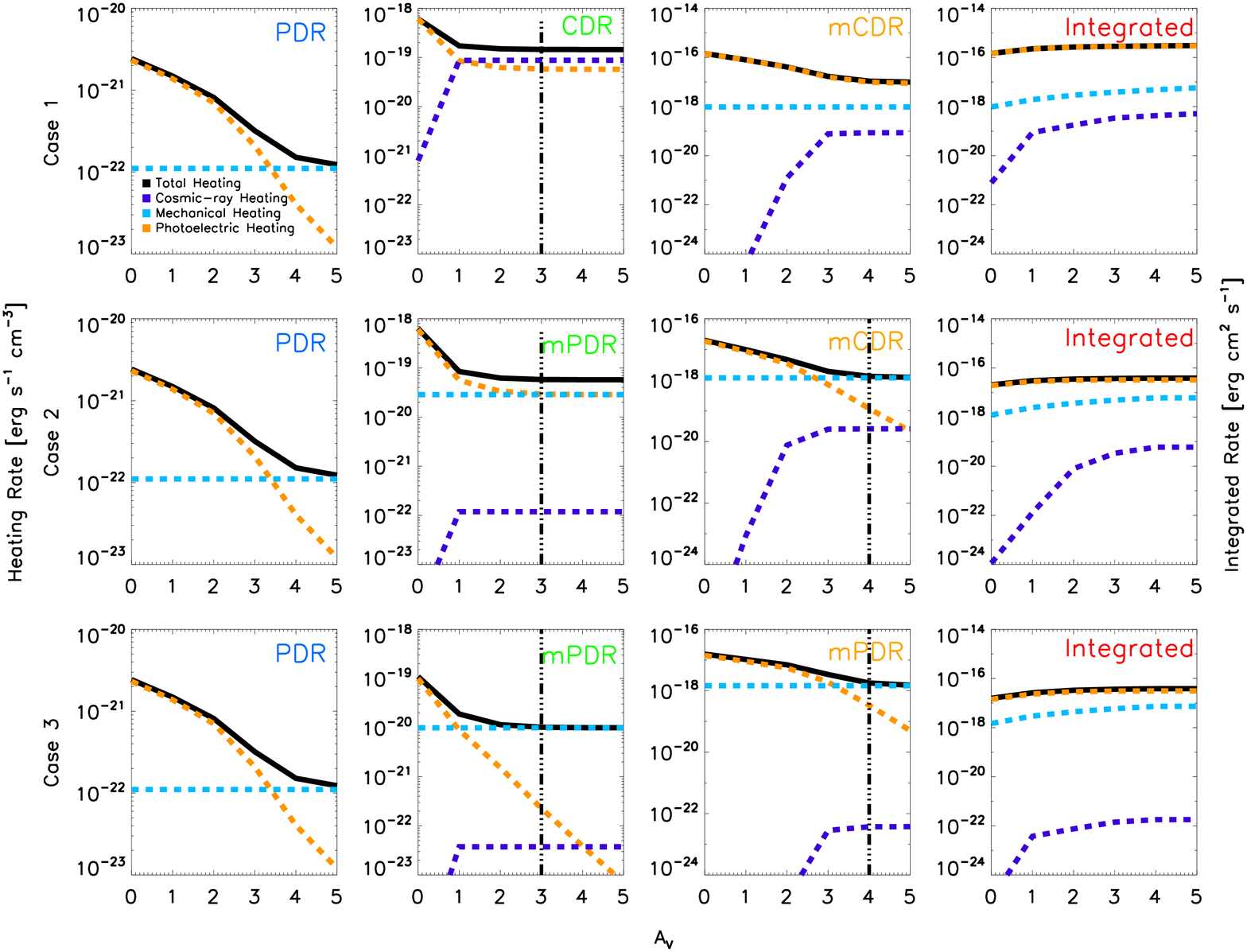}
\caption{Heating rate compared with A$_V$, or depth into cloud.  The total heating is shown in black, the cosmic ray heating is shown in
purple, the mechanical heating in blue, and the photoelectric heating in orange.  The top row is for case 1 with 1 mPDR, 1 CDR, and 1 
mCDR.  The middle row is case 2 with 2 mPDRs and 1mCDR and the bottom row is case 3 with 3 mPDRs.  The first column is for the first component
in each case, the mPDR, and it is the same in all three cases.  The second and third columns represent the second and third ISM phases for each case. The vertical 
dotted line in these plots represents the edge of the PDR, as some of the models we use have A$_V$<5.  Finally,
the last column is the integrated heating rates for all ISM phases.}
\label{fig:heating}
\end{figure*}

Since we have concluded that in NGC 253 mechanical heating plays a key role in the gas excitation, we can compare the total mechanical heating with the other heating 
sources to see which type of heating is dominant.  In order to understand the evolution of heating throughout the cloud, we plot the heating rates as a function of A$_V$,
or depth into the cloud.  We do this for each ISM phase separately as well as for the total heating of the composite system.  We compare all three of our model scenarios: 
case 1 with two
mCDRs, case 2 with one mPDR and one mCDR, and case 3 with two mPDRs.  The photoelectric heating includes all related heating mechanisms such as H$_2$ pumping, CII ionization, 
viscous heating etc.  The heating breakdown is shown in Figure~\ref{fig:heating}.

\subsection{Case 1}
For the first ISM phase, the mPDR, the total heating (in black) is produced mostly by photoelectric heating from the PDR since there is only 1\% mechanical heating.
However, at A$_V$=3.5 the mechanical heating overtakes the photoelectric heating and becomes the dominant heating source.  In all cases, the first PDR phase is kept constant, 
thus for the next sections we will not discuss it further.

For each ISM phase, as well as the integrated heating, at A$_V$=0, or the edge of the cloud, photoelectric heating is the dominant heating source, as expected in a PDR.    
However, in the second ISM phase, the pure CDR, the cosmic ray heating quickly becomes equally, and slightly more important than photoelectric heating at A$_V$=1. 
The third phase, the mCDR has an $\alpha=0.01$, 
and despite a very low mechanical heating rate, the mechanical heating rate is at least 10 times higher than the cosmic ray heating and the photoelectric heating rate 
dominates the heating through 
A$_V$=5.  Looking at the integrated heating,
we see that this case (as well as all cases) is dominated by photoelectric heating. Even though this case includes two ISM phases with an enhanced cosmic ray 
ionization rate, the CR heating never exceeds that of the mechanical heating. 

\subsection{Case 2}
In contrast, the integrated heating for case 2, with two mPDRs and one mCDR, shows although photoelectric heating is the dominant mechanism, mechanical heating still contributes
a significant fraction of the heating and CR heating is insignificant.  We see that for the second ISM phase, 
the emission is dominated by photoelectric heating up to A$_V$=2 and then mechanical heating becomes equally as important. Here cosmic rays are relatively unimportant.  The mCDR model has an enhanced 
cosmic ray ionization rate, and is reflected in the heightened CR heating rate.  However, in order to reproduced the observed HCN and HNC 
fluxes, mechanical heating is still necessary and is consistently more important than the cosmic ray heating rate by a factor of at least 50.  Thus, in case 2, although we included an
enhanced cosmic ray ionization rate in one of the ISM phases, it is unimportant to the total heating of the gas.

\subsection{Case 3}
Finally we study case 3, where there are 3 mPDRs.  In this case, both the second and the third 
phase have an $\alpha =0.1$ and neither has an enhanced cosmic ray ionization rate.  
In both mPDR phases, the photoelectric heating
drops off sharply, and is replaced by mechanical heating at A$_V$=1 and A$_V$=3 for phase 2 and 3 respectively.  Studying the integrated heating, photoelectric
heating still dominates, but the mechanical heating is even more important here, and the CR heating is about 5 orders of magnitude less important.

From these plots, we can conclude that in a PDR with any amount of mechanical heating, mechanical heating is more important the cosmic ray rate,
regardless of if it is enhanced.  However, mechanical heating being the dominant heating source at the edge of a particular cloud does not
mean that it is the overall most efficient source of heating (see the Integrated heating plots).  For example, in case 3 with 3 mPDRs, mechanical 
heating becomes the 
dominant heating source in each individual PDR by at least A$_V$=3.5, yet if we integrate the heating of the composite model we find that mechanical heating is 
only responsible for 20$\%$ of the total heating, while photoelectric heating heats 80$\%$. Similarly, if we integrate the 
contributions from each heating source in the composite model for case 2, we find 84$\%$ photoelectric heating at 16$\%$ mechanical 
heating.  In both cases 1 and 2, cosmic rays play a negligible role in heating the gas, which may be expected since only the third 
phase in case 2 has an enhanced cosmic ray ionization rate.  However, when we integrate the heating phases of case 1, with two
phases with an increased ionization rate, we find 98$\%$ of the heating from photoelectric heating and the remaining 2$\%$ from 
mechanical heating.  The cosmic ray heating contributes only 0.17$\%$ of the total heating.  Therefore, we suggest that although
we cannot rule out cosmic ray contributions to heating the gas, the dominant heating mechanisms are photoelectric heating and mechanical 
heating.  In addition, although mechanical heating may dominate the heating after A$_V$=3 in individual clouds, it still only contributes a maximum of 20$\%$ of the 
heating in the cloud, in agreement with the upper limit prediction from \citet{2013A&A...550A..12R}.  This suggests that the source of feedback 
exciting the high density gas is not directly from star formation, but instead perhaps from the molecular outflows which provide mechanical feedback 
without the production of cosmic rays \citet{2013Natur.499..450B}.

We estimate the mass of each ISM phase and find that regardless of the excitation mechanisms, the mass is relatively well constrained for each case.  The first phase, an mPDR
with 1\% mechanical heating,
has a mass of 6$\times 10^7$ M$_\odot$.  The second phase (either CDR or mPDR) has a mass of 2-5$\times 10^7$ M$_\odot$ and the third phase (either mPDR or mCDR) 
has a mass of 2-4$\times 10^6$ M$_\odot$.  With the exception of the third phase, the masses of each ISM phase are all approximately the same.  In addition, the total 
mass of each system is well conserved.  We find a mass of 8-11$\times10^7$ M$_\odot$.  This mass is in agreement with $3\times10^8$ M$_\odot$ in an 80'' 
beam \citep{2008A&A...490...77W} since if the gas were evenly distributed, we would expect a mass of 5$\times10^7$M$_\odot$.  The gas mass is concentrated in the
center of the galaxy, thus it is reasonable to expect a mass a few times higher than the evenly distributed mass. In all cases the third phase contributes at least 
60\% of the CO emission but
represents a maximum of 15\% of the mass.  
Also, in the third ISM phase of all cases, the CO column density is less than 1\% of the total CO column.  This suggests 
that the CO is mainly dissociated in this phase, but still responsible for $\sim 60\%$ of the C.

NGC 253 is an ideal case in which we have any observation desired. It is a nearby, well-studied, star-forming galaxy that gives us a representative 
view of environments undergoing rapid star formation.  We have identified that in these galaxies, although mechanical heating is important, heating by UV photons
is still the dominant heating source.  We have also determined that in this environment, cosmic rays at best play a minor role in heating the molecular gas. In this sense, 
any star forming galaxy within a similar luminosity class as NGC 253, would have similar contributions of heating mechanisms. Yet in LIRGs and ULIRGs, there are many 
other factors in play, including AGN which produce x rays, mergers, and powerful outflows, which increase turbulent heating.  It is conceivable that mechanical heating is the domninant source
of excitation is such galaxies, or that X-rays are much more important.

For example, studies of Mrk 231 show that the ISM is being processed both by star formation and supernovae \citep{2014A&A...561A..27G, 2010A&A...518L..41F}.  NGC 6240
on the other hand, shows clear evidence of shocks and mechanical heating \citep{2013ApJ...762L..16M}.  In NGC 1068, \citet{2012ApJ...755...57H} find that the gas 
is either excited by x rays or shocks, but cannot differentiate the two.  These galaxies are some examples of (U)LIRGs that show the complexity of disentangling heating
mechanisms in these luminous systems. They also demonstrate that in the case of (U)LIRGs it is rarely only UV photons that contribute to heating the molecular gas.  Although
NGC 253 is not a LIRG, the starburst in the center represents a similar environment, and already demonstrates the need to include an additional heating mechanisms.

\section{Conclusions}
\label{sec:conc}
We observe the starburst nucleus of NGC 253 with Herschel SPIRE in order to understand the heating and excitation of the gas.
\begin{enumerate}
\item We extract the fluxes of all the lines in our spectra using FTFitter, and correct them for the changing beam size of 
SPIRE.  We do a basic LVG analysis, using both $^{12}$CO and $^{13}$CO in order to constrain the parameter space. By mapping the $\chi^2$ values for the full parameter space, we can get a handle on the degeneracies and limit the parameter space further.  
\item Using the limits set by the LVG analysis, we perform a PDR analysis.  We use not only the $^{12}$ and $^{13}$CO, but include the
HCN and HNC ladders as well, to constrain the dense gas.  Although we get a good fit with 3 PDR models, we can see that the 
the solutions are not reasonable and an additional heating source is required.  The additional heating source is needed because we cannot reproduce a high enough temperature with just a PDR model to match the observed CO emission.

\item Using PDR models with a constant mechanical heating rate (mPDRs) and PDR models with both a constant mechanical heating rate and a cosmic
ray ionization rate equal to 750 times our galactic value (mCDRs), we can fit $^{12}$CO, $^{13}$CO, HCN, HNC.  We fit the four molecular 
ladders for three different cases, case 1 is 1 mPDR with 2 mCDRs, case 2 is 2 mPDRs and 1 mCDR, and case 3 is 3 mPDRs.  
\item We find that all cases provide a good fit.  We cannot fit all four molecules with only enhanced cosmic ray models, we require the first 
model to be mechanically heated. Although we require mechanical heating to fit the observed transitions, we cannot rule out the existence of cosmic 
ray heating in addition to the mechanical heating.

\item Taking these three different cases of fits, we study the heating balance as a function of depth into the cloud.  We find that in 
case 1, with an $\alpha$ of just 1\% mechanical heating dominates cosmic ray heating at all depths of the cloud.  In both cases 2 and 3, mechanical heating 
quickly becomes the dominant heating mechanism as the cloud depth increases for individual models.  In case 3, we see a similar picture with mechanical heating
quickly dominating the photoelectric heating in individual models.
\item Regardless of mechanical heating dominating the heating sources, in our most heavily mechanically 
heated case (case 3), it only contributes about 20$\%$ of the total heating, while photoelectric heating is still the dominant overall
heating source.  In all cases, including case 1 with an enhanced ionization rate for two of the ISM phases, cosmic ray heating is 
negligible to the total heating of the cloud, as seen in the integrated heating rates of Figure~\ref{fig:heating}. 
\item We estimate the mass of each ISM phase and regardless of case, the mass is well constrained.  The first phase, excited by predominantly photoelectric 
heating, has a mass of 6$\times10^7$ M$_\odot$.  The second phase has a mass of 3$\times10^7$ M$_\odot$ and the third phase has a mass of 9$\times10^6$ M$_\odot$.  Although
the phases have similar masses, the third phase is responsible for $\sim 60\%$ of the CO emission.

\end{enumerate}
\bibliographystyle{aa}
\bibliography{bib_file.bib}
\begin{acknowledgements}

We would like to thank Edward Polehampton for his help preparing the SPIRE observations.  SPIRE has been developed by a consortium of institutes led by
Cardiff Univ. (UK) and including: Univ. Lethbridge (Canada);
NAOC (China); CEA, LAM (France); IFSI, Univ. Padua (Italy);
IAC (Spain); Stockholm Observatory (Sweden); Imperial College London, RAL, UCL-MSSL, UKATC, Univ. Sussex (UK);
and Caltech, JPL, NHSC, Univ. Colorado (USA). This development has been supported by national funding agencies:
CSA (Canada); NAOC (China); CEA, CNES, CNRS (France);
ASI (Italy); MCINN (Spain); SNSB (Sweden); STFC, UKSA
(UK); and NASA (USA). The Herschel spacecraft was designed, built, tested, and 
launched under a contract to ESA managed by the Herschel/Planck Project team by 
an industrial consortium under the overall responsibility of the prime contractor 
Thales Alenia Space (Cannes), and including Astrium (Friedrichshafen) responsible 
for the payload module and for system testing at spacecraft level, Thales Alenia 
Space (Turin) responsible for the service module, and Astrium (Toulouse) responsible 
for the telescope, with in excess of a hundred subcontractors. HCSS / HSpot / HIPE is 
a joint development (are joint developments) by the Herschel Science Ground 
Segment Consortium, consisting of ESA, the NASA Herschel Science Center, and 
the HIFI, PACS and SPIRE consortia.

\end{acknowledgements}

\end{document}